\documentclass[preprint,showpacs,aps,amsmath,amssymb]{revtex4}
\usepackage{graphicx}

\newcommand{\bra}[1]{\langle {#1} |}
\newcommand{\ket}[1]{| {#1} \rangle}

\begin{document}

\title{Efficient calculation of the quasiparticle random-phase approximation
 matrix}

\author{Paolo Avogadro}
\affiliation{%
Department of Physics \& Astronomy, Texas A\&M
University-Commerce, Commerce, Texas 75428, USA}
\affiliation{%
 RIKEN Nishina Center, Wako, 351-0198, Japan
}%
\author{Takashi Nakatsukasa}%
\affiliation{%
 RIKEN Nishina Center, Wako, 351-0198, Japan
}%
\affiliation{%
 Center for Computational Sciences, University of Tsukuba,
Tsukuba, 305-8571, Japan
}%

\date{\today}

\begin{abstract}
We present an efficient numerical technique to evaluate the matrix
of the (quasiparticle)-random-phase approximation,
using the finite amplitude method (FAM).
The method is tested in calculation of monopole excitations in $^{120}$Sn,
compared with result obtained with the former iterative FAM.
The neutron-pair-transfer modes are calculated with the
present method and their character change in neutron-rich Pb isotopes
is discussed.
Computational aspects of different FAM approaches are also discussed
for future applications to a large-scale computation.
\end{abstract}


\maketitle


\section{Introduction}
\label{intro}
The random-phase approximation (RPA) is a leading theory in studies of
elementary excitations in nuclei and other quantum many-body systems.
The RPA, which is equivalent to the small-amplitude limit of the
time-dependent mean-field theory, is commonly known in the matrix
form \cite{RS80,BR86} as
\begin{equation}
\label{RPA_eq}
 \begin{pmatrix}
  A     &    B  \\
 -B^{*} &   -A^{*}
 \end{pmatrix}
\begin{pmatrix}
 X^{(n)}    \\
 Y^{(n)}
\end{pmatrix}
=
\omega_n
\begin{pmatrix}
 X^{(n)}    \\
 Y^{(n)}
\end{pmatrix}
.
\end{equation}
Here, $X^n$ and $Y^n$ are respectively called forward and backward amplitudes
of the $n$-th RPA normal modes.
We refer to the matrix in the left-hand side of Eq. (\ref{RPA_eq})
as the RPA matrix, or as the quasiparticle RPA (QRPA) matrix in the case
that the mean fields contain the pair potential.
In nuclear mean-field models, the evaluation of the (Q)RPA matrix is
a tedious task and requires significant efforts for programing the
computer code.
The main purpose of the present paper is to present a feasible and efficient
numerical method to evaluate the (Q)RPA matrix.

The finite amplitude method (FAM) was proposed as a feasible numerical approach
to the calculation of the strength functions \cite{NIY07}.
Recently, it has been extended to the quasiparticle-basis
representation to include the pairing correlations \cite{AN11}.
In this method, instead of calculating the RPA normal modes,
one solves the linear response equation with an external field at
a given frequency.
The FAM has been successfully applied to the Skyrme energy
functionals in different representations;
the three-dimensional (3D) coordinate-mesh representation with no symmetry
restriction but no pairing \cite{INY09,INY11}, the quasiparticle
basis with the radial-mesh representation for spherical nuclei \cite{AN11},
and that with the harmonic-oscillator-basis representation for
axially deformed nuclei \cite{Sto11}.
The FAM is also shown to be superior to the conventional approach,
with respect to numerical costs \cite{NIY07,INY09,Sto11}.

So far, the FAM has been utilized for the calculation of the strength
functions, using iterative algorithms.
Hereafter, this approach is referred to as ``iterative FAM'' (i-FAM).
However, when we are interested in low-lying discrete modes of excitation,
it is desirable to obtain the RPA normal modes in Eq. (\ref{RPA_eq}).
In this paper, we show another usage of the FAM, for explicit
construction of the RPA matrix.
We call this approach ``matrix FAM'' (m-FAM) in this paper.
The method only requires a straightforward extension of the former FAM
for the strength function.
Since the complicated programing is required for calculation of
the RPA matrix for realistic energy functionals,
this new method is useful for the verification of the existing/developing
computer codes of the (Q)RPA, as well.

The paper is organized as follows:
In Sec. \ref{sec: FAM_RPA}, after a brief introduction to
the finite amplitude method, we propose a new method, m-FAM, of calculating
the RPA matrix for systems without pairing correlations.
In Sec. \ref{sec: FAM_QRPA}, we present a method applicable to
superfluid systems, namely the m-FAM for the QRPA.
Numerical results are shown in Sec.~\ref{sec: numerical} for
monopole strength function in $^{120}$Sn and the neutron-pair transfer
to excited $0^+$ states in neutron-rich Pb isotopes.
Comparisons between i-FAM and m-FAM are shown in Sec.~\ref{sec: comparison},
in terms of the computational point of view.
Finally, the summary is given in Sec.~\ref{sec: conclusion}.
%
%
\section{FAM calculation of the RPA matrix}
\label{sec: FAM_RPA}

First, let us discuss the case without the pairing correlation.
It is customary to use 
the canonical single-particle representation, which we adopt in this paper too.
Thus, the forward and backward amplitudes in Eq. (\ref{RPA_eq}) have
particle-hole (ph) indices, $(X_{ph},Y_{ph})$,
in which the particle and hole states are assumed to be
eigenstates of the single-particle Hamiltonian at the ground states:
$h[\rho_0] \ket{\phi_k} =\epsilon_k\ket{\phi_k}$.
We use the notation $\phi_h$ for hole orbitals ($h=1,\cdots,A$)
and $\phi_p$ for particle orbitals ($p=A+1,\cdots,\infty$).
The matrices $A$ and $B$ in Eq. (\ref{RPA_eq}),
which have two pairs of ph indices, are given by
\begin{equation}
\label{AB}
A_{ph,p'h'}= (\epsilon_p-\epsilon_h) \delta_{pp'}\delta_{hh'}
    + \frac{\partial h_{ph}}{\partial \rho_{p'h'}} , \quad\quad
B_{ph,p'h'}=
    \frac{\partial h_{ph}}{\partial \rho_{h'p'}} .
\end{equation}
The residual interactions, $\partial h/\partial \rho$,
are the derivatives of the
single-particle Hamiltonian $h[\rho]$ with respect to one-body density,
evaluated at the ground-state density ($\rho=\rho_0$).
The explicit evaluation of these residual interactions is
the most demanding part in the RPA calculations, with respect both to
the computational cost and to the programming task.

\subsection{The Finite Amplitude Method (FAM)}
\label{sec:FAM}

In this subsection, we recapitulate the FAM \cite{NIY07}.
The upper part in the left-hand side of Eq. (\ref{RPA_eq}) leads to
\begin{eqnarray}
&& \sum_{p'h'} \left( A_{ph,p'h'}X_{p'h'} + B_{ph,p'h'}Y_{p'h'} \right)
= (\epsilon_p-\epsilon_h) X_{ph} + \delta h_{ph} , \\
\label{delta_h_ph}
&& \delta h_{ph} \equiv \sum_{p'h'} \left(
 \frac{\partial h_{ph}}{\partial \rho_{p'h'}} X_{p'h'}
 + \frac{\partial h_{ph}}{\partial \rho_{h'p'}} Y_{p'h'} \right) .
\end{eqnarray}
In Ref. \cite{NIY07}, we have proposed the FAM which provides an easy way
to evaluate $\delta h_{ph}$ for a given vector $(X,Y)$.
All we need to do is to calculate the single-particle Hamiltonian $h[\rho]$
at the density slightly different from the ground state $\rho=\rho_0$
as follows.
Provided that a real parameter $\eta$ is small enough to allow us to neglect
the $\eta^2$ and higher-order terms,
\begin{eqnarray}
\label{delta_h}
\delta h &=& \frac{1}{\eta}
\left( h[\rho_\eta] - h[\rho_0] \right) , \\
\label{rho_eta}
\rho_\eta &\equiv& \rho_0+\eta \delta\rho
= \sum_h \ket{\psi_h} \bra{\bar\psi_h} + O(\eta^2),
\end{eqnarray}
where the hole orbitals are slightly modified from the ground-state
canonical states $\phi_h$, in different manners between the ket and bra states,
as follows:
\begin{equation}
\label{psi_h}
\ket{\psi_h}\equiv \ket{\phi_h} +\eta  \sum_p X_{ph} \ket{\phi_p} ,\quad
\bra{\bar\psi_h}\equiv \bra{\phi_h} + \eta \sum_p Y_{ph} \bra{\phi_p} .
\end{equation}
Note that the use of different bra's and ket's in Eq. (\ref{rho_eta})
leads to non-hermitian $\delta\rho$ \cite{NIY07}.
From Eq. (\ref{rho_eta}), one can easily see that this is necessary to
obtain the well-known relation in the RPA;
$\delta\rho_{ph}=\eta X_{ph}$ and $\delta\rho_{hp}=\eta Y_{ph}$.

Now, using the finite-difference calculation of Eq. (\ref{delta_h}),
we can evaluate $\delta h_{ph}$ for a given $(X,Y)$.
In earlier works on calculations of the strength functions
\cite{NIY07,INY09,INY11},
the linear response equation was solved with
an iterative algorithm starting from an arbitrary initial vector $(X,Y)$.
The convergence of the iteration provides
the self-consistent RPA amplitudes $(X,Y)$.
In this iterative process, the RPA matrix elements themselves,
$A$ and $B$, are never calculated,
instead, only the product of the RPA matrix and the vector, such as
$AX+BY$, are calculated.

\subsection{Calculation of the RPA matrix}
\label{sec:FAM_RPA_matrix}

Now, we present the essential idea of the present paper, that is
a method of calculating the RPA matrix without
explicit evaluation of the derivative, $\partial h/\partial\rho$.
The idea is very simple and
immediately understood from Eq. (\ref{delta_h_ph}).
Namely, if we choose the vector as
\begin{equation}
\label{XY=0}
X_{mi}=\delta_{mp'}\delta_{ih'} ,\quad Y_{mi}=0 ,
\end{equation}
where $m>A$ and $i\leq A$,
then, the FAM calculation of $\delta h_{ph}$ in Eq. (\ref{delta_h})
leads to $\partial h_{ph}/\partial\rho_{p'h'}$.
If we choose
\begin{equation}
\label{YX=0}
X_{mi}=0, \quad Y_{mi}=\delta_{mp'}\delta_{ih'} ,
\end{equation}
then, it produces $\partial h_{ph}/\partial\rho_{h'p'}$.
Since the FAM can provide the vector $(AX+BY)_{ph}$ for a given $(X,Y)$,
we can obtain the matrix elements,
$A_{ph,p'h'}$ and $B_{ph,p'h'}$,
by choosing the vectors, Eqs. (\ref{XY=0}) and (\ref{YX=0}), respectively.
Therefore, the RPA matrix can be explicitly constructed by the
calculation of the single-particle Hamiltonian $h[\rho_\eta]$ only,
using the FAM.

Sometimes, we resort to a different from of the RPA equation.
In such cases, a difference choice of the vector may be convenient.
For instance,
the RPA equation (\ref{RPA_eq}) can be recast into the canonical form
in terms of the normal-mode coordinate $Q^{(n)}$ and momentum $P^{(n)}$
\cite{RS80}:
\begin{equation}
\label{RPA_eq_QP}
(A+B)(A-B)Q^{(n)} =\omega_n^2 Q^{(n)} , \quad
(A-B)(A+B)P^{(n)} =\omega_n^2 P^{(n)} .
\end{equation}
Mapping these normal coordinates on given collective variables,
we may obtain the RPA (Thouless-Valatin) collective mass,
which has been recently used to study the collective quadrupole dynamics
\cite{HSNMM10,SH11,HSYNMM11}.
The numerical solution of the RPA eigenvalue problem can be simplified
by further
transforming Eq. (\ref{RPA_eq_QP}) into its hermitian form \cite{RS80}.
An advantage of this approach in the large-scale parallel computing
has been recently demonstrated as well \cite{YN11}.
These calculations do not require the matrix $A$ and $B$ separately, but
need the sum of them, $(A\pm B)_{ph,p'h'}$.
This is directly accessible with the FAM choosing $(X,Y)$ as follows:
\begin{equation}
X_{mi}=\delta_{mp'}\delta_{ih'}, \quad Y_{mi}=\pm\delta_{mp'}\delta_{ih'} .
\end{equation}

To construct the full RPA matrix, the residual fields should be
calculated, according to Eq. (\ref{delta_h}),
for all the independent unit vectors, Eqs. (\ref{XY=0}) and (\ref{YX=0}).
However, the present m-FAM does not resort to iterative solver.
Thus, if the dimension of the RPA matrix is small,
the m-FAM is computationally more efficient than the i-FAM.
The detailed discussion about the computational aspects is shown
in Sec.~\ref{sec: comparison}.

\section{FAM calculation of the QRPA matrix}
\label{sec: FAM_QRPA}

In this section, the result in Sec.~\ref{sec: FAM_RPA} is generalized
for the QRPA with the pairing correlations.
The mean-field Hamiltonian in the ground state
is diagonalized in the quasiparticle states:
\begin{equation}
H=\sum_{kl} \left(
(h_{kl}-\lambda\delta_{kl}) c_k^\dagger c_l
+ \frac{1}{2} \Delta_{kl} c_k^\dagger c_l^\dagger
+ \frac{1}{2} \Delta_{kl}^* c_l c_k
\right)
=E_0+\sum_\mu E_\mu a_\mu^\dagger a_\mu .
\end{equation}
Here, $(c_k,c_k^\dagger)$ are the annihilation and creation operators of
a particle at the basis state $k$, and $(a_\mu,a_\mu^\dagger)$ are
those of the quasiparticle states.
The single-particle (ph) Hamiltonian
$h_{kl}[\rho,\kappa,\kappa^*]$ and
the pair (pp,hh) potential $\Delta_{kl}[\rho,\kappa,\kappa^*]$
are now functionals of the one-body density $\rho_{kl}$ and
the pair tensors $(\kappa_{kl},\kappa^*_{kl})$.
The forward and backward amplitudes have two-quasiparticle (2qp) indices,
$X_{\mu\nu}$ and $Y_{\mu\nu}$, and
the QRPA matrix has a pair of 2qp indices,
$A_{\mu\nu,\mu'\nu'}$ and $B_{\mu\nu,\mu'\nu'}$.
Here, the quasiparticle states $(U_{k\mu},V_{k\mu})$
are chosen to be states corresponding to
the positive quasiparticle energies $E_\mu$.

A detailed formalism of the FAM for the QRPA is found in Ref.~\cite{AN11}.
Here, we briefly summarize the main result.
Again, the upper part in the left-hand side of Eq. (\ref{RPA_eq})
is written as \cite{AN11}
\begin{eqnarray}
\label{AX+BY_QRPA}
&&\sum_{\mu'<\nu'} \left( A_{\mu\nu,\mu'\nu'} X_{\mu'\nu'}
+ B_{\mu\nu,\mu'\nu'} Y_{\mu'\nu'} \right)
= (E_\mu + E_\nu) X_{\mu\nu} + \delta H_{\mu\nu}^{(20)} , \\
&& \delta H^{20}_{\mu\nu}=
  \left(
  U^{\dagger} \delta h             V^{*}
- V^{\dagger} \delta \Delta^{(-)*} V^{*} 
+ U^{\dagger} \delta \Delta^{(+)}  U^{*}
- V^{\dagger} \delta h^T           U^{*} \right)_{\mu\nu} ,
\end{eqnarray}
where the induced fields, $\delta h_{kl}$ and $\delta\Delta^{(\pm)}_{kl}$,
can be calculated with a small parameter $\eta$ as \cite{AN11}
\begin{eqnarray}
\delta h(\omega) &=&
\frac{1}{\eta} \left(
h\left[\rho_\eta,\kappa_\eta^{(+)},\kappa_\eta^{(-)*} \right]
- h\left[\rho_0,\kappa_0,\kappa_0^* \right]
\right)
 , \\
\delta \Delta^{(+)}(\omega) &=&
\frac{1}{\eta} \left(
\Delta\left[\rho_\eta,\kappa_\eta^{(+)},\kappa_\eta^{(-)*} \right]
- \Delta\left[\rho_0,\kappa_0,\kappa_0^* \right]
\right)
 , \\
\delta \Delta^{(-)}(\omega) &=&
\frac{1}{\eta} \left(
\Delta\left[\rho_\eta^\dagger,\kappa_\eta^{(-)},\kappa_\eta^{(+)*} \right]
- \Delta\left[\rho_0,\kappa_0,\kappa_0^* \right]
\right)
 .
\end{eqnarray}
Here, the FAM densities, $\rho_\eta$ and $\kappa_\eta^{(\pm)}$,
are slightly changed from those at the ground state and
calculated with the modified quasiparticle wave functions as
\begin{equation}
\label{rho_kappa_eta}
\begin{split}
\rho_\eta
&\equiv (V^* + \eta UX)(V + \eta U^* Y)^T \\
\kappa_\eta^{(+)}
&\equiv (V^* + \eta UX)(U + \eta V^* Y)^T \\
\kappa_\eta^{(-)}
&\equiv (V^* + \eta UY^*)(U + \eta V^* X^*)^T .
\end{split}
\end{equation}
From these FAM formulae,
we may calculate $\delta H_{\mu\nu}^{(20)}$ for a given vector $(X,Y)$,
without the explicit calculation of the complicated residual interactions,
$v_{\mu\nu,\mu'\nu'}$.

Now, we apply the same trick as we did in Sec.~\ref{sec:FAM_RPA_matrix},
to calculate the QRPA matrix elements.
To obtain $A_{\mu\nu,\mu'\nu'}$, we choose
\begin{equation}
\label{vecA}
X_{\alpha\beta}=\delta_{\alpha\mu'}\delta_{\beta\nu'} , \quad
Y_{\alpha\beta}=0 ,
\end{equation}
while $B_{\mu\nu,\mu'\nu'}$ is obtained by choosing
\begin{equation}
\label{vecB}
X_{\alpha\beta}=0 . \quad
Y_{\alpha\beta}=\delta_{\alpha\mu'}\delta_{\beta\nu'} .
\end{equation}
If we want $(A\pm B)$, we may use the following
\begin{equation}
X_{\alpha\beta}=\delta_{\alpha\mu'}\delta_{\beta\nu'} . \quad
Y_{\alpha\beta}=\pm\delta_{\alpha\mu'}\delta_{\beta\nu'} , \quad
\rightarrow\quad
(A\pm B)_{\mu\nu,\mu'\nu'} .
\end{equation}
In this way, the QRPA matrix can be explicitly constructed
by the FAM approach.

\section{Numerical results}
\label{sec: numerical}

In this section, we show the numerical results of the QRPA based on
the HFB ground state.
The present FAM approach to the calculation of the QRPA matrix has been
implemented in the computer code {\sc hfbrad}
for the spherical Hartree-Fock-Bogoliubov (HFB) calculation \cite{HFBRAD}.
The static HFB solution is obtained in the radial coordinate space
discretized with a mesh of 0.1 fm in the box of 20 fm.
The maximum angular momenta for the neutrons and protons are set to be
$j_{\rm max}=21/2$ and $15/2$, respectively.
We use the energy density functional of the SkM* parameter set and
of the volume-type pairing
\begin{equation}
V_{\rm pair}(\vec{r}-\vec{r}') = V_0
(1-P_\sigma) \delta(\vec{r}-\vec{r}') ,
\end{equation}
with the strength $V_0=-90$ MeV fm$^3$.
The quasiparticle energy cutoff is set at $E_{\rm qp}^c=60$ MeV,
unless otherwise specified.
Then, the FAM construction of the QRPA matrix is performed.
We include all the two-quasiparticle states in the quasiparticle
space truncated by $E_{\rm qp}^c$.
The number of two-quasiparticle states with the $J^\pi=0^+$ is
$N_{\rm 2qp}=1741$ for $E_{\rm qp}^c=60$ MeV for $^{120}$Sn.

\subsection{Monopole strength function}

We calculate the QRPA matrix using the FAM presented
in Sec.~\ref{sec: FAM_QRPA}.
To obtain the QRPA matrix of $A$,
we repeat the FAM calculation adopting vectors
$(X_{\alpha\beta},Y_{\alpha\beta})=(\delta_{\alpha\mu'}\delta_{\beta\nu'},0)$,
with $N_{\rm 2qp}$ different pairs of $\mu'\nu'$.
Then, to obtain the matrix $B$,
we use another $N_{\rm 2qp}$ kinds of vectors,
$(X_{\alpha\beta},Y_{\alpha\beta})=(0,\delta_{\alpha\mu'}\delta_{\beta\nu'})$.
Once the $A$ and $B$ matrix is calculated,
we resort to a routine {\sc zgeev} in the Lapack libraries \cite{lapack}
to diagonalize the QRPA matrix in Eq. (\ref{RPA_eq}).
The diagonalization of the QRPA matrix produces the normal-mode
excitation energies $\omega_n$ and eigenvectors $(X^{(n)},Y^{(n)})$.
This defines the $n$-th normal-mode creation operator,
\begin{equation}
\Omega_n^\dagger = \sum_{\mu>\nu} \left(
X^{(n)}_{\mu\nu} a_\mu^\dagger a_\nu^\dagger
-Y^{(n)}_{\mu\nu} a_\nu a_\mu \right) .
\end{equation}
Then, we calculate the transition matrix elements of a one-body
operator $F$ between the ground state $|0\rangle$ and
the $n$-th excited state $|n\rangle = \Omega_n^\dagger |0\rangle$, as
\begin{equation}
\label{transition_strength}
 \langle n| F |0\rangle =
 \langle {\rm HFB}| [\Omega_n, F] |{\rm HFB}\rangle = \sum_{\mu>\nu} \left(
X^{(n)*}_{\mu\nu} F_{\mu\nu}^{20} 
+Y^{(n)*}_{\mu\nu} F_{\mu\nu}^{02} 
\right)
\end{equation}
$F^{20}$ and $F^{02}$ are $a^\dagger a^\dagger$- and $aa$-parts
of the one-body operator $F$ defined by
\begin{equation}
F=\sum_{kl} F_{kl} c_k^\dagger c_l
=F_0+\sum_{\mu>\nu} \left( F_{\mu\nu}^{20} a_\mu^\dagger a_\nu^\dagger
                       +F_{\mu\nu}^{02} a_\nu a_\mu \right) 
 +\sum_{\mu\nu} F_{\mu\nu}^{11} a_\mu^\dagger a_\nu .
\end{equation}
In case that
the operator $F$ is a hermitian operator conserving the particle number,
they are given by
\begin{equation}
F^{20}_{\mu\nu}= \left(
U^\dagger F V^* - V^\dagger F^T U^* \right)_{\mu\nu} ,
\end{equation}
and $F^{02}=F^{20*}$.

\begin{figure}[t]
 \centering
 \includegraphics[width=10cm]{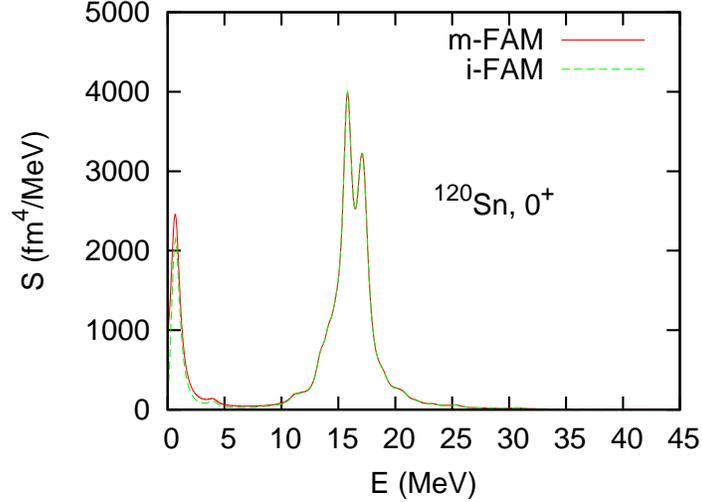}
 \caption{(Color online)
Isoscalar monopole strength function for $^{120}$Sn
calculated with m-FAM and i-FAM.
The green dotted curve is obtained with the iterative method
without explicit calculation of the QRPA matrix,
while the red curve is the present approach to the QRPA matrix
with a Lorentzian smearing of $\gamma=1$ MeV.
}
 \label{fig:sn120}
 \end{figure}

The result of the present approach to
the isoscalar monopole strength function ($F\equiv \sum_k r_k^2$)
for $^{120}$Sn is shown in Fig. \ref{fig:sn120}.
In this nucleus, the protons are in the normal phase, while
the neutrons are in the paired superfluid phase with an average
gap of 1.3 MeV.
We also perform the iterative solution of the
linear-response calculation (i-FAM) for a given energy $E$,
following the procedure in Ref.~\cite{AN11}.
We adopt an iterative algorithm of the generalized conjugate residual (GCR)
method \cite{Saad03}.
The energy $E$ is taken from 0 to 45 MeV with $\Delta E=0.1$ MeV, to which
we add an imaginary part $\gamma/2$, $E\rightarrow E+i\gamma/2$.
In this calculation, we adopt $\gamma=1$ MeV.
To compare the results,
we smear the transition amplitudes of Eq.~(\ref{transition_strength})
in the m-FAM, with a Lorentzian function of
\begin{equation}
\label{strength_function}
 S(\omega;F) = \frac{\gamma/2}{\pi} \sum_n
 \frac{|\langle n| F |0\rangle|^2}{(\omega-\omega_n)^{2}+\gamma^{2}/4} .
\end{equation}

It is clear from Fig. \ref{fig:sn120} that the two independent calculations
give essentially the identical result.
There is a small deviation in the peak strength near
zero energy,
which is due to the fact that, in the vicinity of the zero energy,
the strength function obtained by
the linear-response calculation with
the complex energy $E\rightarrow E+i\gamma/2$ is slightly different
from the one using
the smoothing with the Lorentzian form of
Eq. (\ref{strength_function}).
Actually, this peak is associated with the Nambu-Goldstone mode of
the pairing rotation for neutrons.
The energy of this spurious mode in $^{120}$Sn
is calculated as $\omega_{\rm NG}=662$ keV.
In principle, we should obtain $\omega_{\rm NG}=0$ because the present
calculation is fully self-consistent.
However, the energy of the spurious mode is extremely sensitive
to numerical errors in calculation of the QRPA matrix elements,
which leads to the sizable energy shift.
Note that other physical excitations are practically not affected by
these errors.
In fact, we have confirmed that this single state carries 99.9 \%\ of
the total strength associated with the neutron number operator,
$|\bra{n}\hat{N}\ket{0}|^2$.

In the i-FAM, we calculate the response function for the
external field of the isoscalar monopole, however, we do not know
the eigenenergies and eigenvectors of the QRPA normal modes.
The present method provides us with this missing information.

\subsection{Pair transfer modes in neutron-rich Pb isotopes}

A feature of the m-FAM different from the i-FAM is
an explicit calculation of the normal mode.
This is useful for studies of low-lying excited states.
In this respect, the m-FAM has an advantage over the i-FAM.
In this section, we study the neutron-pair-transfer strength for
low-lying excited $0^+$ states in Pb isotopes.
The pair-transfer operators for the $(L,S)=(0,0)$ pair are defined as
\begin{eqnarray}
P^\dagger&\equiv& \frac{1}{\sqrt{4\pi}} \int d\vec{r} f(r)
 \psi_n^\dagger(\vec{r}\downarrow) \psi_n^\dagger(\vec{r}\uparrow), \\
P&\equiv& \frac{1}{\sqrt{4\pi}} \int d\vec{r} f(r)
 \psi_n(\vec{r}\uparrow) \psi_n(\vec{r}\downarrow),
\end{eqnarray}
for pair-addition and removal operators, respectively.
Here, $\psi_n^\dagger(\vec{r},\sigma)$ and $\psi_n(\vec{r},\sigma)$ 
indicate the creation and annihilation field operators, respectively,
for neutrons with spin $\sigma$ at the position $\vec{r}$.
The radial form factor $f(r)$ is chosen as unity in this paper.
We calculate the transition probabilities from the ground state
of the nucleus with the neutron number $N$
to an $J^\pi=0^+$ excited state $|n\rangle$ in the nucleus with $N\pm 2$,
\begin{eqnarray}
B({\rm add};{\rm gs}\rightarrow n)&\equiv&
 |\langle n | P^\dagger | 0 \rangle |^2 ,\\
B({\rm rem};{\rm gs}\rightarrow n)&\equiv&
 |\langle n | P | 0 \rangle |^2 .
\end{eqnarray}
They are given by exactly the same expression as
 Eq. (\ref{transition_strength}).
For the pair addition mode,
$F^{20}$ and $F^{02}$ are replaced by
\begin{eqnarray}
(P^\dagger)^{20}_{\mu\nu}&=&\frac{1}{\sqrt{4\pi}}\sum_{k>0} \left(
U^*_{k\mu} U^*_{\bar{k}\nu} - U^*_{\bar{k}\mu} U^*_{k\nu} \right) ,\\
(P^\dagger)^{02}_{\mu\nu}&=& \frac{-1}{\sqrt4{\pi}}\sum_{k>0} \left(
V_{k\mu} V_{\bar{k}\nu} - V_{\bar{k}\mu} V_{k\nu} \right) ,
\end{eqnarray}
respectively.
For the pair removal mode $P$, we have the same expressions by
interchanging $U\leftrightarrow V$ with an opposite sign. 

\begin{table}[t]
\caption{Calculated chemical potentials $\lambda_n$ [ MeV ]
and average pairing gaps $\Delta_n$ [ MeV ] for neutrons in Pb isotopes.
The protons are in the normal phase, $\Delta_p=0$, for all isotopes.
}
\label{tab:Pb_HFB}
     \begin{center}
        \begin{tabular}{c c c}
        \hline
        $N$ &  $\lambda_n$ & $\Delta_n$  \\
        \hline
        126    & \ \  $-6.27$\ \ \  & $0.00$ \\
        128    &  $-4.87$  & $0.77$ \\
        130    &  $-4.68$  & $1.03$ \\
        132    &  $-4.50$  & $1.19$ \\
        134    &  $-4.33$  & $1.31$ \\
        136    &  $-4.17$  & $1.42$ \\
        138    &  $-4.03$  & $1.53$ \\
        140    &  $-3.90$  & $1.62$ \\
        142    &  $-3.80$  & $1.70$ \\
        \hline
        \end{tabular}
     \end{center}
\end{table}

\begin{figure}
 \centerline{
 \includegraphics[height=0.8\textheight]{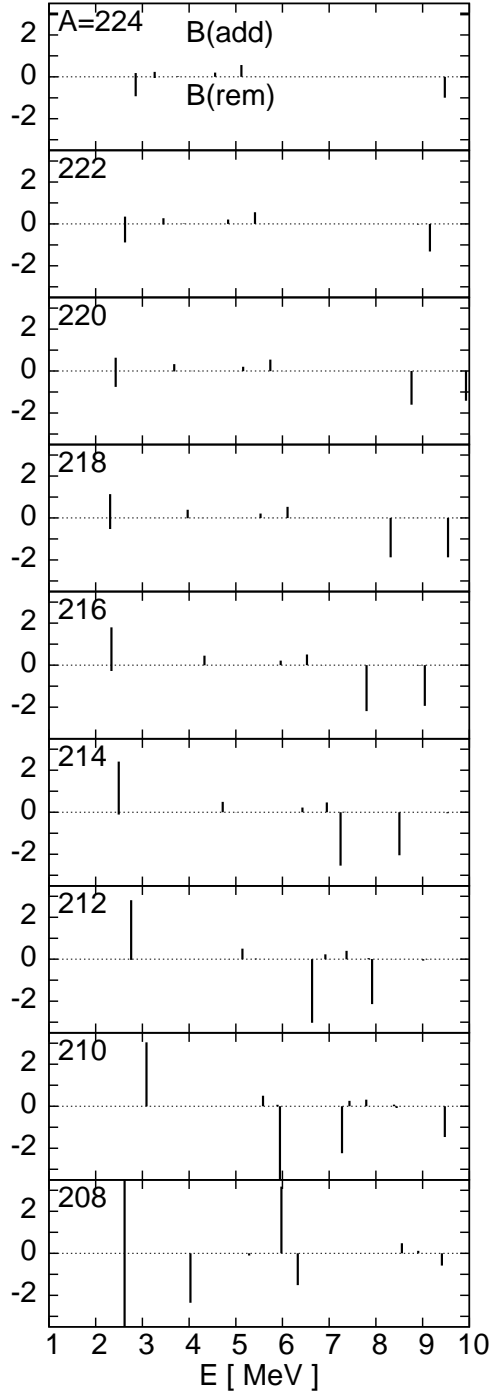}
}
 \caption{Neutron pair transfer strengths
to excited states in Pb isotopes.
The energy is regarded as the approximate excitation energy in the
nucleus with $N\pm 2$, except for $^{208}$Pb.
The pair-addition strength is shown as the positive value,
$B({\rm add};N_{\rm gs} \rightarrow (N+2)_{\rm ex})$,
while
the removal strength is shown as the negative strength,
$-B({\rm rem};N_{\rm gs} \rightarrow (N-2)_{\rm ex})$.
These quantities are dimensionless,
since the radial form factor is chosen as unity, $f(r)=1$.
Note that the following pair strengths are out of vertical range:
$B({\rm add})=7.47$ and $B({\rm rem})=4.93$ at $E=2.61$ MeV in $^{208}$Pb
and $B({\rm rem})=3.74$ at $E=5.95$ MeV in $^{210}$Pb.
}
  \label{fig:transfer_Pb}
 \end{figure}

We perform the calculation for even-even Pb isotopes from $N=126$ to
$N=142$.
The HFB calculation with the SkM* parameters predicts the
ground-state properties in Table~\ref{tab:Pb_HFB}.
The neutron pairing gap gradually increases as the neutron number.
In Fig.~\ref{fig:transfer_Pb}, we show two-neutron-transfer
strengths as a function of the (Q)RPA normal-mode excitation energy $E$.
The ground state in $^{208}$Pb is in the normal phase with
$\Delta_n=\Delta_p=0$, due
to its doubly closed-shell configuration.
In this nucleus,
the lowest mode around $E=2.6$ MeV corresponds to the ground state
in neighboring nuclei with $N\pm 2$.
This corresponds to the pairing vibration
which carries significant strengths both for
two-neutron addition and removal modes;
$B({\rm add};^{208}{\rm Pb}\rightarrow^{210}{\rm Pb})=7.47$ and
$B({\rm rem};^{208}{\rm Pb}\rightarrow^{206}{\rm Pb})=4.93$.
For the removal mode, the calculation suggests another $0^+$ state
with a sizable strength, located at about 1.4 MeV higher than the
ground state in $^{206}$Pb.
The excited $0^+$ state at $E_{\rm ex}=1.17$ MeV
was observed by the $(p,t)$ reaction
whose cross section is about 10 \% of that of the ground state
\cite{Lan77,Tak83}.

In the normal phase, the chemical potential $\lambda$ is not uniquely
determined.
Although the pair strengths are invariant,
the energies of the normal modes $(E^{\rm add},E^{\rm rem})$
depend on the choice of the chemical potential
in $^{208}$Pb.
However, the sum of them, $E^{\rm add}+E^{\rm rem}$,
does not depend on the chemical potential.
In fact, the chemical potential for $^{208}$Pb
in Table~\ref{tab:Pb_HFB} is determined by the condition that the
pair addition and removal modes have the same excitation energy.

For superfluid isotopes with the finite pairing gap ($\Delta_n\neq 0$),
the energy in Fig.~\ref{fig:transfer_Pb} can be regarded as
\begin{equation}
E=E_{\rm ex}(N\pm 2) - E_{\rm gs}(N) \mp 2\lambda
 \approx E_{\rm ex}(N\pm 2) - E_{\rm gs}(N\pm 2) .
\end{equation}
There are no significant pair-removal strength
in excited $0^+$ states for $^{210-216}$Pb.
The lowest state is calculated to be about 6 MeV higher
than the ground state.
In contrast, the pair-addition strength is present at low energy;
$E_{\rm ex}=3.0$ MeV for $^{210}$Pb and the energy is even lowered by
increasing the neutron number, leading to the minimum value of
$E_{\rm ex}=2.3$ MeV in $^{218}$Pb.
Further increasing the neutron number, this lowest $0^+$ transfer mode
gradually changes from the pair-addition into the pair-removal character.
In $^{224}$Pb, the lowest mode carries very little pair-addition
strength.

\section{Comparison in computational point of view}
\label{sec: comparison}

\subsection{Iterative FAM (i-FAM)}

The iterative algorithms adopted in the former i-FAM calculations
\cite{NIY07,INY09,INY11,AN11,Sto11} are slightly different from each other.
Nevertheless, these calculations are composed of common ingredients:
\begin{enumerate}
\item The (Q)RPA matrix is not explicitly constructed.
\item The linear-response equation is solved for a fixed energy
$\omega=E+i\gamma/2$ with a given external field.
\end{enumerate}
Thus, the i-FAM solves
a linear algebraic system of the form $M\vec{x}=\vec{b}$.
The algorithm \cite{Saad03}
involves matrix-by-vector multiplication, $M\vec{p}$
for a given $\vec{p}$,
to produce a succession of vectors converging to the solution:
\begin{equation}
\begin{split}
 \vec{x}_0,~ \vec{x}_1, \cdots, \vec{x}_n &\rightarrow \vec{x} \\ 
 \vec{r}_n\equiv \vec{b}-M\vec{x}_n &\rightarrow 0 .
\end{split}
\end{equation}
The precision of the solution can be measured by 
$\epsilon_{n} = || \vec{r}_n ||^2/||\vec{b}||^2$ and
the iterative procedure stops when $\epsilon_n < \epsilon$.
With this technique, we do not need to compute the values of the
matrix elements of $M$, since it is enough to know the 
product of $M\vec{p}$ for a given vector $\vec{p}$.
The GCR algorithm requires calculation of $M\vec{p}$
twice at each iteration.
From the view point of the memory resources,
the method has a great advantage because the biggest arrays to be used
in the code are of the order of $2N_{\rm 2qp}$,
instead of those for the matrix, $2N_{\rm 2qp}\times 2N_{\rm 2qp}$.

For the present calculation for $^{120}$Sn,
we have employed the following parameters:
the precision $\epsilon =10^{-5}$,
the FAM parameter in Eq. (\ref{rho_kappa_eta}) $\eta=10^{-8}$,
the energy range of $0<E<45$ MeV discretized with a mesh of $\Delta E=0.1$ MeV
with a smoothing parameter $\gamma=1$ MeV.
The speed of the calculation depends on the number of iterations required
to reach the convergence, which is particularly affected by the precision
parameter.
In the present case, the number of iterations ranges from about 30
at low energy, to about 160 where the strength has a peak (around 15-20 MeV).
To start the iteration procedure, we need an initial vector $\vec{x}_0$.
We used the solution obtained at $E-\Delta E$ as the initial vector
for the energy $E$.

\subsection{Matrix FAM (m-FAM)}

To construct the full QRPA matrix, we need to compute the right-hand side
of Eq. (\ref{AX+BY_QRPA}) for $2 N_{\rm 2qp}$ kinds of vectors $(X,Y)$.
In case that $N_{\rm 2qp}$ is relatively small,
this is the most time-consuming part for the m-FAM.
It also requires
the memory capacity to store the matrix of order of
$2N_{\rm 2qp}\times 2N_{\rm 2qp}$.
However, there is a practical advantage in the m-FAM; the calculation of
the residual fields, $\delta H_{\mu\nu}^{20}$, is easier in the m-FAM
than the i-FAM.
For a given vector $(X,Y)$, we calculate the densities,
$(\rho_\eta,\kappa^{(\pm)}_\eta)$, in Eq. (\ref{rho_kappa_eta}).
In the i-FAM, since the vector $(X,Y)$ is updated every iteration,
we construct $\rho_\eta$ by the matrix operation as
\begin{equation}
\label{drhokappa_original}
(\rho_\eta)_{ij} = \sum_{\alpha}
\left(V^*_{i\alpha} + \eta \sum_\beta U_{i\beta} X_{\beta\alpha}\right)
\left(V_{j\alpha} + \eta \sum_\gamma U^*_{j\gamma} Y_{\gamma\alpha}\right) ,
\end{equation}
and we have similar expressions for $\kappa^{(\pm)}$.
Here, we need to sum over the three quasiparticle indices
 $(\alpha,\beta,\gamma)$.
In the case of m-FAM, we can omit these summations,
because of the simple form of $(X,Y)$ such as
Eqs. (\ref{vecA}) and (\ref{vecB}).
For instance, the calculation of $A_{\mu\nu,\mu'\nu'}$ can be carried out
by adopting Eq. (\ref{vecA}).
Then, the densities are simply given by
\begin{equation}
\begin{split}
(\rho_\eta)_{ij} &= \rho_0 + \eta U_{i\mu'} V_{j\nu'} ,\\
(\kappa_\eta^{(+)})_{ij} &= \kappa_0 + \eta U_{i\mu'} U_{j\nu'} ,\\
(\kappa_\eta^{(-)})_{ij} &= \kappa_0 + \eta V^*_{i\nu'}V^*_{j\mu'} .
\end{split}
\end{equation}
Thus, the calculation of these densities is faster in the m-FAM than
in the i-FAM.
In the present numerical calculation, we have found that this
reduces the computation time for building the QRPA matrix, becoming
1/5 of the original time with Eq. (\ref{drhokappa_original}).

After the calculation of the QRPA matrix, we diagonalize the
matrix using the Lapack libraries.
The computational task required for this diagonalization scales
as $(2N_{\rm 2qp})^3$,
while that for the calculation of the QRPA matrix elements scales as
as $(2N_{\rm 2qp})^2$.
Thus, increasing the quasiparticle model space,
the diagonalization will eventually become a hot spot in the computation.

\subsection{Comparison of computational time for $^{120}$Sn}

In Table~\ref{tab:convergence},
we show the relative CPU time of the calculation of monopole strength
in i-FAM and the QRPA calculation with the m-FAM, increasing the
number of two-quasiparticle states (increasing $E_{\rm qp}^c$).
In the present calculation with the quasiparticle-energy cutoff of
$E_{\rm qp}^c=60$ MeV,
the m-FAM is six times faster than the i-FAM.
The computation time for i-FAM shows a weak scaling with respect to
the size of the model space,
close to the linear dependence on $N_{\rm 2qp}$.
On the other hand, the computation time for m-FAM indicates a scaling
between $N_{\rm 2qp}^2$ and $N_{\rm 2qp}^3$.
Therefore, for treating the larger model space,
the i-FAM has a computational advantage over the m-FAM,
with respect to this scaling property and the memory requirement.

\begin{table}[t]
\caption{Computational time of
the iterative FAM (i-FAM) for the monopole strength
function and the matrix FAM (m-FAM) for all the $J^\pi=0^+$ eigenstates
with different cutoff energies ($E_{\rm qp}^c$),
relative to that of i-FAM with $E_{\rm qp}^c=60$ MeV.
The size of the QRPA matrix is determined by the number of $0^+$
two-quasiparticle states ($N_{\rm 2qp}$) in the model space,
which is varied by changing $E_{\rm qp}^c$.
}
     \begin{center}
        \begin{tabular}{r c r r}
        \hline   
        \multicolumn{1}{c}{$E_{\rm qp}^c$} &  $2\times N_{\rm 2qp}$\ \  &  i-FAM &  m-FAM  \\
        \hline
        60 MeV     &      3482   &    1   &      0.16  \\
        80 MeV     &      4656   &    1.43   &      0.38  \\
        100 MeV    &      5842   &    1.93   &      0.60  \\
        120 MeV    &      7156   &   2.64   &     1.26  \\
        140 MeV    &      8336   &   3.27   &     1.77  \\ 
        160 MeV    &      9528   &   4.08   &     2.56  \\     
        \hline 
        \end{tabular}
     \end{center}
\label{tab:convergence}
\end{table}

\subsection{Parallelization}

In contrast to the present FAM calculations for spherical nuclei,
the FAM calculation for deformed systems may require significant
computational resources with modern massively parallel supercomputers.
In the case of the i-FAM,
the calculation of the strength functions
at a given energy is independent from the other energies.
This leads to an obvious kind of parallelization with the number of
processors equal to that of energy points.
Typically, the number of energy points is order of 100 at most.
Since the iteration process cannot be parallerized,
the parallel computation with processors beyond this number is not trivial.

In contrast, the large-scale parallel computation can be easily done in m-FAM. 
The m-FAM calculates the matrices,
$A_{\mu\nu,\mu'\nu'}$ and $B_{\mu\nu,\mu'\nu'}$.
For a given pair of $\mu'\nu'$, the matrix elements of $A_{\mu\nu,\mu'\nu'}$
are produced from the vector of Eq. (\ref{vecA}), and
those of $B_{\mu\nu,\mu'\nu'}$ are from that of Eq. (\ref{vecB}).
Since these procedures are completely independent,
it is easy to utilize the processors whose number is
as large as $2\times N_{\rm 2qp}$.
For deformed systems, the number of two-quasiparticle states could be
order of $10^5$, even assuming the axial symmetry \cite{TE10,YN11}.
Thus, the massive parallelization of this magnitude can be 
achieved in the m-FAM.

\section{Conclusions}\label{sec: conclusion}
We have presented a feasible method to construct the matrix of
the RPA and the QRPA,
based on the idea of the finite amplitude method (FAM).
Since all the residual interactions are numerically estimated by the FAM,
it does not require complicated programming.
The method can be easily implemented with existing HF and HFB codes,
to turn them into the RPA and QRPA codes.
We have used the {\sc hfbrad} code to test the present method.
An advantage of the present method, called matrix FAM (m-FAM),
over the i-FAM is that the m-FAM provides the normal-mode
vectors explicitly.

In the computational aspect for a large-scale problem
(increasing $N_{\rm 2qp}$),
a disadvantage of the m-FAM over the i-FAM is
that the computational task scales as $(N_{\rm 2qp})^2$ for the
calculation of the matrix elements and as $(N_{\rm 2qp})^3$ for
the diagonalization.
In addition, it requires a large memory capacity to store the
QRPA matrix of $2N_{\rm 2qp}\times 2N_{\rm 2qp}$.
On the other hand, there are some advantages as well.
For problems of small dimensions, the computational time is shorter
than the former iterative FAM (i-FAM).
For problems of large dimensions, the m-FAM may resort to a
massively parallel computer, because the calculation of the matrix
elements can be easily parallelized up to the number of processors
equal to twice of the number of two-quasiparticle states.
Thus, the m-FAM could be a strong candidate of the new QRPA code
for deformed superfluid nuclei, for the use of the massively
parallel computers.

In many respects, the i-FAM and the m-FAM have complimentary character
to each other.
In addition, there are intriguing developments in the iterative
approaches to the low-lying normal-mode solutions:
The conjugate gradient method was adopted in Refs.~\cite{Muta02,IH03}
for the RPA solutions in the 3D real-space representation.
Very recently, the iterative Arnoldi method, developed in Ref.~\cite{Toi10},
has been applied to the calculation of the lowest solutions of the QRPA in
spherical nuclei \cite{CTP12}.
These novel technologies and the present FAM may provide
a powerful tool
for developments of the efficient and feasible (Q)RPA codes for
deformed systems.

\section*{Acknowledgments}
This work is supported by Grant-in-Aid for Scientific Research(B)
(No. 21340073) and on Innovative Areas (No. 20105003).
It is also supported by SPIRE, MEXT, Japan.
P.A. was supported by the U.S. Department of Energy under
Contract No. DE-FG02-08ER41533, DE-SC000497, and
DE-FG02-06ER41407 (JUSTIPEN), during completion of this work.
The numerical calculations were performed in part
on RIKEN Integrated Cluster of Clusters (RICC).

\bibliography{myself,nuclear_physics,current}

\begin{thebibliography}{17}
\expandafter\ifx\csname natexlab\endcsname\relax\def\natexlab#1{#1}\fi
\expandafter\ifx\csname bibnamefont\endcsname\relax
  \def\bibnamefont#1{#1}\fi
\expandafter\ifx\csname bibfnamefont\endcsname\relax
  \def\bibfnamefont#1{#1}\fi
\expandafter\ifx\csname citenamefont\endcsname\relax
  \def\citenamefont#1{#1}\fi
\expandafter\ifx\csname url\endcsname\relax
  \def\url#1{\texttt{#1}}\fi
\expandafter\ifx\csname urlprefix\endcsname\relax\def\urlprefix{URL }\fi
\providecommand{\bibinfo}[2]{#2}
\providecommand{\eprint}[2][]{\url{#2}}

\bibitem[{\citenamefont{Ring and Schuck}(1980)}]{RS80}
\bibinfo{author}{\bibfnamefont{P.}~\bibnamefont{Ring}} \bibnamefont{and}
  \bibinfo{author}{\bibfnamefont{P.}~\bibnamefont{Schuck}},
  \emph{\bibinfo{title}{The nuclear many-body problems}},
  (\bibinfo{publisher}{Springer-Verlag}, \bibinfo{address}{New
  York}, \bibinfo{year}{1980}).

\bibitem[{\citenamefont{Blaizot and Ripka}(1986)}]{BR86}
\bibinfo{author}{\bibfnamefont{J.-P.} \bibnamefont{Blaizot}} \bibnamefont{and}
  \bibinfo{author}{\bibfnamefont{G.}~\bibnamefont{Ripka}},
  \emph{\bibinfo{title}{Quantum Theory of Finite Systems}}
  (\bibinfo{publisher}{MIT Press}, \bibinfo{address}{Cambridge},
  \bibinfo{year}{1986}).

\bibitem[{\citenamefont{Nakatsukasa et~al.}(2007)\citenamefont{Nakatsukasa,
  Inakura, and Yabana}}]{NIY07}
\bibinfo{author}{\bibfnamefont{T.}~\bibnamefont{Nakatsukasa}},
  \bibinfo{author}{\bibfnamefont{T.}~\bibnamefont{Inakura}}, \bibnamefont{and}
  \bibinfo{author}{\bibfnamefont{K.}~\bibnamefont{Yabana}},
  \bibinfo{journal}{Phys. Rev. C} \textbf{\bibinfo{volume}{76}},
  \bibinfo{pages}{024318} (\bibinfo{year}{2007}).

\bibitem[{\citenamefont{Avogadro and Nakatsukasa}(2011)}]{AN11}
\bibinfo{author}{\bibfnamefont{P.}~\bibnamefont{Avogadro}} \bibnamefont{and}
  \bibinfo{author}{\bibfnamefont{T.}~\bibnamefont{Nakatsukasa}},
  \bibinfo{journal}{Phys. Rev. C} \textbf{\bibinfo{volume}{84}},
  \bibinfo{pages}{014314} (\bibinfo{year}{2011}).

\bibitem[{\citenamefont{Inakura et~al.}(2009)\citenamefont{Inakura,
  Nakatsukasa, and Yabana}}]{INY09}
\bibinfo{author}{\bibfnamefont{T.}~\bibnamefont{Inakura}},
  \bibinfo{author}{\bibfnamefont{T.}~\bibnamefont{Nakatsukasa}},
  \bibnamefont{and} \bibinfo{author}{\bibfnamefont{K.}~\bibnamefont{Yabana}},
  \bibinfo{journal}{Phys. Rev. C} \textbf{\bibinfo{volume}{80}},
  \bibinfo{pages}{044301} (\bibinfo{year}{2009}).

\bibitem[{\citenamefont{Inakura et~al.}(2011)\citenamefont{Inakura,
  Nakatsukasa, and Yabana}}]{INY11}
\bibinfo{author}{\bibfnamefont{T.}~\bibnamefont{Inakura}},
  \bibinfo{author}{\bibfnamefont{T.}~\bibnamefont{Nakatsukasa}},
  \bibnamefont{and} \bibinfo{author}{\bibfnamefont{K.}~\bibnamefont{Yabana}},
  \bibinfo{journal}{Phys. Rev. C} \textbf{\bibinfo{volume}{84}},
  \bibinfo{pages}{021302} (\bibinfo{year}{2011}).

\bibitem[{\citenamefont{Stoitsov et~al.}(2011)\citenamefont{Stoitsov,
  Kortelainen, Nakatsukasa, Losa, and Nazarewicz}}]{Sto11}
\bibinfo{author}{\bibfnamefont{M.}~\bibnamefont{Stoitsov}},
  \bibinfo{author}{\bibfnamefont{M.}~\bibnamefont{Kortelainen}},
  \bibinfo{author}{\bibfnamefont{T.}~\bibnamefont{Nakatsukasa}},
  \bibinfo{author}{\bibfnamefont{C.}~\bibnamefont{Losa}}, \bibnamefont{and}
  \bibinfo{author}{\bibfnamefont{W.}~\bibnamefont{Nazarewicz}},
  \bibinfo{journal}{Phys. Rev. C} \textbf{\bibinfo{volume}{84}},
  \bibinfo{pages}{041305} (\bibinfo{year}{2011}).

\bibitem[{\citenamefont{Hinohara et~al.}(2010)\citenamefont{Hinohara, Sato,
  Nakatsukasa, Matsuo, and Matsuyanagi}}]{HSNMM10}
\bibinfo{author}{\bibfnamefont{N.}~\bibnamefont{Hinohara}},
  \bibinfo{author}{\bibfnamefont{K.}~\bibnamefont{Sato}},
  \bibinfo{author}{\bibfnamefont{T.}~\bibnamefont{Nakatsukasa}},
  \bibinfo{author}{\bibfnamefont{M.}~\bibnamefont{Matsuo}}, \bibnamefont{and}
  \bibinfo{author}{\bibfnamefont{K.}~\bibnamefont{Matsuyanagi}},
  \bibinfo{journal}{Phys. Rev. C} \textbf{\bibinfo{volume}{82}},
  \bibinfo{pages}{064313} (\bibinfo{year}{2010}).

\bibitem[{\citenamefont{Sato and Hinohara}(2011)}]{SH11}
\bibinfo{author}{\bibfnamefont{K.}~\bibnamefont{Sato}} \bibnamefont{and}
  \bibinfo{author}{\bibfnamefont{N.}~\bibnamefont{Hinohara}},
  \bibinfo{journal}{Nuclear Physics A} \textbf{\bibinfo{volume}{849}},
  \bibinfo{pages}{53 } (\bibinfo{year}{2011}).

\bibitem[{\citenamefont{Hinohara et~al.}(2011)\citenamefont{Hinohara, Sato,
  Yoshida, Nakatsukasa, Matsuo, and Matsuyanagi}}]{HSYNMM11}
\bibinfo{author}{\bibfnamefont{N.}~\bibnamefont{Hinohara}},
  \bibinfo{author}{\bibfnamefont{K.}~\bibnamefont{Sato}},
  \bibinfo{author}{\bibfnamefont{K.}~\bibnamefont{Yoshida}},
  \bibinfo{author}{\bibfnamefont{T.}~\bibnamefont{Nakatsukasa}},
  \bibinfo{author}{\bibfnamefont{M.}~\bibnamefont{Matsuo}}, \bibnamefont{and}
  \bibinfo{author}{\bibfnamefont{K.}~\bibnamefont{Matsuyanagi}},
  \bibinfo{journal}{Phys. Rev. C} \textbf{\bibinfo{volume}{84}},
  \bibinfo{pages}{061302} (\bibinfo{year}{2011}).

\bibitem[{\citenamefont{Yoshida and Nakatsukasa}(2011)}]{YN11}
\bibinfo{author}{\bibfnamefont{K.}~\bibnamefont{Yoshida}} \bibnamefont{and}
  \bibinfo{author}{\bibfnamefont{T.}~\bibnamefont{Nakatsukasa}},
  \bibinfo{journal}{Phys. Rev. C} \textbf{\bibinfo{volume}{83}},
  \bibinfo{pages}{021304} (\bibinfo{year}{2011}).

\bibitem[{\citenamefont{Bennaceur and Dobaczewski}(2005)}]{HFBRAD}
\bibinfo{author}{\bibfnamefont{K.}~\bibnamefont{Bennaceur}} \bibnamefont{and}
  \bibinfo{author}{\bibfnamefont{J.}~\bibnamefont{Dobaczewski}},
  \bibinfo{journal}{Computer Physics Communications}
  \textbf{\bibinfo{volume}{168}}, \bibinfo{pages}{96 } (\bibinfo{year}{2005}).

\bibitem[{\citenamefont{Anderson et~al.}(1999)\citenamefont{Anderson, Bai,
  Bischof, Blackford, Demmel, Dongarra, Croz, Greenbaum, Hammarling, McKenney
  et~al.}}]{lapack}
\bibinfo{author}{\bibfnamefont{E.}~\bibnamefont{Anderson}},
  \bibinfo{author}{\bibfnamefont{Z.}~\bibnamefont{Bai}},
  \bibinfo{author}{\bibfnamefont{C.}~\bibnamefont{Bischof}},
  \bibinfo{author}{\bibfnamefont{S.}~\bibnamefont{Blackford}},
  \bibinfo{author}{\bibfnamefont{J.}~\bibnamefont{Demmel}},
  \bibinfo{author}{\bibfnamefont{J.}~\bibnamefont{Dongarra}},
  \bibinfo{author}{\bibfnamefont{J.~D.} \bibnamefont{Croz}},
  \bibinfo{author}{\bibfnamefont{A.}~\bibnamefont{Greenbaum}},
  \bibinfo{author}{\bibfnamefont{S.}~\bibnamefont{Hammarling}},
  \bibinfo{author}{\bibfnamefont{A.}~\bibnamefont{McKenney}},
  \bibnamefont{et~al.}, \emph{\bibinfo{title}{LAPACK Users' Guide}}
  (\bibinfo{publisher}{Society for Industrial and Applied Mathematics},
  \bibinfo{year}{1999}).

\bibitem[{\citenamefont{Saad}(2003)}]{Saad03}
\bibinfo{author}{\bibfnamefont{Y.}~\bibnamefont{Saad}},
  \emph{\bibinfo{title}{Iterative methods for sparse linear systems}}
  (\bibinfo{publisher}{SIAM}, \bibinfo{address}{Philadelphia},
  \bibinfo{year}{2003}).

\bibitem[{\citenamefont{Lanford}(1977)}]{Lan77}
\bibinfo{author}{\bibfnamefont{W.~A.} \bibnamefont{Lanford}},
  \bibinfo{journal}{Phys. Rev. C} \textbf{\bibinfo{volume}{16}},
  \bibinfo{pages}{988} (\bibinfo{year}{1977}).

\bibitem[{\citenamefont{Takahashi et~al.}(1983)\citenamefont{Takahashi,
  Murakami, Morita, Orihara, Ishizaki, and Yamaguchi}}]{Tak83}
\bibinfo{author}{\bibfnamefont{M.}~\bibnamefont{Takahashi}},
  \bibinfo{author}{\bibfnamefont{T.}~\bibnamefont{Murakami}},
  \bibinfo{author}{\bibfnamefont{S.}~\bibnamefont{Morita}},
  \bibinfo{author}{\bibfnamefont{H.}~\bibnamefont{Orihara}},
  \bibinfo{author}{\bibfnamefont{Y.}~\bibnamefont{Ishizaki}}, \bibnamefont{and}
  \bibinfo{author}{\bibfnamefont{H.}~\bibnamefont{Yamaguchi}},
  \bibinfo{journal}{Phys. Rev. C} \textbf{\bibinfo{volume}{27}},
  \bibinfo{pages}{1454} (\bibinfo{year}{1983}).

\bibitem[{\citenamefont{Terasaki and Engel}(2010)}]{TE10}
\bibinfo{author}{\bibfnamefont{J.}~\bibnamefont{Terasaki}} \bibnamefont{and}
  \bibinfo{author}{\bibfnamefont{J.}~\bibnamefont{Engel}},
  \bibinfo{journal}{Phys. Rev. C} \textbf{\bibinfo{volume}{82}},
  \bibinfo{pages}{034326} (\bibinfo{year}{2010}).

\bibitem[{\citenamefont{Muta et~al.}(2002)\citenamefont{Muta, Iwata, Hashimoto,
  and Yabana}}]{Muta02}
\bibinfo{author}{\bibfnamefont{A.}~\bibnamefont{Muta}},
  \bibinfo{author}{\bibfnamefont{J.-I.} \bibnamefont{Iwata}},
  \bibinfo{author}{\bibfnamefont{Y.}~\bibnamefont{Hashimoto}},
  \bibnamefont{and} \bibinfo{author}{\bibfnamefont{K.}~\bibnamefont{Yabana}},
  \bibinfo{journal}{Prog. Theor. Phys.} \textbf{\bibinfo{volume}{108}},
  \bibinfo{pages}{1065} (\bibinfo{year}{2002}).

\bibitem[{\citenamefont{Imagawa and Hashimoto}(2003)}]{IH03}
\bibinfo{author}{\bibfnamefont{H.}~\bibnamefont{Imagawa}} \bibnamefont{and}
  \bibinfo{author}{\bibfnamefont{Y.}~\bibnamefont{Hashimoto}},
  \bibinfo{journal}{Phys. Rev. C} \textbf{\bibinfo{volume}{67}},
  \bibinfo{pages}{037302} (\bibinfo{year}{2003}).

\bibitem[{\citenamefont{Toivanen et~al.}(2010)\citenamefont{Toivanen, Carlsson,
  Dobaczewski, Mizuyama, Rodr\'iguez-Guzm\'an, Toivanen, and Vesel\'y}}]{Toi10}
\bibinfo{author}{\bibfnamefont{J.}~\bibnamefont{Toivanen}},
  \bibinfo{author}{\bibfnamefont{B.~G.} \bibnamefont{Carlsson}},
  \bibinfo{author}{\bibfnamefont{J.}~\bibnamefont{Dobaczewski}},
  \bibinfo{author}{\bibfnamefont{K.}~\bibnamefont{Mizuyama}},
  \bibinfo{author}{\bibfnamefont{R.~R.} \bibnamefont{Rodr\'iguez-Guzm\'an}},
  \bibinfo{author}{\bibfnamefont{P.}~\bibnamefont{Toivanen}}, \bibnamefont{and}
  \bibinfo{author}{\bibfnamefont{P.}~\bibnamefont{Vesel\'y}},
  \bibinfo{journal}{Phys. Rev. C} \textbf{\bibinfo{volume}{81}},
  \bibinfo{pages}{034312} (\bibinfo{year}{2010}).

\bibitem[{\citenamefont{Carlsson et~al.}(2012)\citenamefont{Carlsson, Toivanen,
  and Pastore}}]{CTP12}
\bibinfo{author}{\bibfnamefont{B.~G.} \bibnamefont{Carlsson}},
  \bibinfo{author}{\bibfnamefont{J.}~\bibnamefont{Toivanen}}, \bibnamefont{and}
  \bibinfo{author}{\bibfnamefont{A.}~\bibnamefont{Pastore}},
  \bibinfo{journal}{Phys. Rev. C} \textbf{\bibinfo{volume}{86}},
  \bibinfo{pages}{014307} (\bibinfo{year}{2012}).
\end{thebibliography}

\end{document}